\documentclass[prl,aps,twocolumn,superscriptaddress,showpacs]{revtex4}
\usepackage{epsfig}
\usepackage{amsmath}

\begin{document}

\title {Heat transport in proximity structures}

\author{E.~V.~Bezuglyi}

\affiliation{Institute for Low Temperature Physics and Engineering,
Kharkov 61103, Ukraine}

\affiliation{Argonne National Laboratory, Argonne IL 60439, U.S.A.}

\author{V.~Vinokur}

\affiliation{Argonne National Laboratory, Argonne IL 60439, U.S.A.}

\date{\today}

\begin{abstract}

We study heat and charge transport through a normal diffusive wire
coupled with a superconducting wire over the region smaller than the
coherence length. Due to partial Andreev reflection of quasiparticles
from the interface, the subgap thermal flow is essentially suppressed
and approaches zero along with energy, which is specific for
diffusive structures. Whereas the electric conductance shows
conventional reentrance effect, the thermal conductance $\kappa$
rapidly decreases with temperature which qualitatively explains the
results of recent experiments. In the Andreev interferometer
geometry, $\kappa$ experiences full-scale oscillations with the order
parameter phase difference.

\pacs{74.25.Fy, 74.45.+c, 73.23.-b}
\end{abstract}

\maketitle

\begin{figure}[tb]
\epsfxsize=8.5cm\epsffile{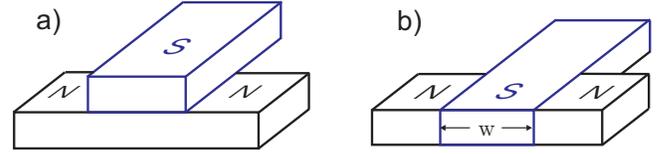}
\caption{Two models of the NS interface: sandwich geometry (a), and
inset geometry (b).} \label{fig1}
\end{figure}

Manifestations of the proximity effect in the electron transport of
the hybrid normal metal-superconductor (NS) structures are in the
focus of current extensive research. Up to now the electric
conductance in NS-hybrids has been receiving much more attention than
thermal conductance; the imbalance can be ascribed to difficulties
one encounters in carrying out thermal transport experiments in
mesoscopic samples. A remarkable breakthrough in recent measurements
\cite{Chandr1,Chandr2} of both the thermal conductance and
thermopower in an $Au$ (N) diffusive wire of the micron length and
submicron cross-size and thickness, where an $Al$ (S) needle-like
sample of similar parameters is deposited across it, as shown in
Fig.~\ref{fig1},a, calls for and motivates a detailed theoretical
investigation of heat and charge transport in mesoscopic NS-hybrids.
In our Letter we develop a theory of thermal and electric
conductances in mesoscopic proximity structures in the diffusive
limit, which, as we show below, describes the experimental situation.
We find that while the electric conductance exhibits the conventional
reentrance behavior, the thermal conductance rapidly decreases with
temperature, as $\kappa\sim T^4$, in a qualitative agreement with
experimental findings.

Experiments \cite{Chandr1,Chandr2} showed a tiny (within a few
percents) change in the electric conductance of $Au$ due to the
proximity effect in accordance with the past studies.  The thermal
conductance, in a contrast, dropped with temperature decreasing, by
the order of magnitude as compared to its value $\kappa_N(T)$ in the
normal state.  This result seems to excellently follow the original
idea by Andreev \cite{And} first applied to the thermal conductance
of the intermediate state of a superconductor \cite{Zavar}: If the
interface resistance is negligibly small, and the wire thickness is
smaller than the coherence length $\xi_0$, then the sandwich-like NS
contact in Fig.~\ref{fig1},a can be modelled by an inset of the
superconductor into the N-wire, Fig.~\ref{fig1},b
\onlinecite{Chandr2}. Then the superconductor lead in
Fig.~\ref{fig1},b plays the role of a quantum barrier for normal
quasiparticles with the energies $E$ smaller than the superconducting
order parameter $\Delta$. Low-energy electrons hit the barrier and
convert into the retro-reflected holes (Andreev reflection), carrying
the same energy back in the opposite direction. Thus, the subgap
thermal flow through the superconductor is blocked, and only
quasiparticles with energies $E>\Delta$ participate in the heat
transport.  At low temperatures, $T \ll \Delta$, the thermal
conductance follows nearly exponential temperature dependence,
$\kappa \sim T\exp(-\Delta/T)$ \cite{And}, used in Ref.\
\onlinecite{Chandr2} for data fitting.

\begin{figure}[tb]
\epsfxsize=8.5cm\epsffile{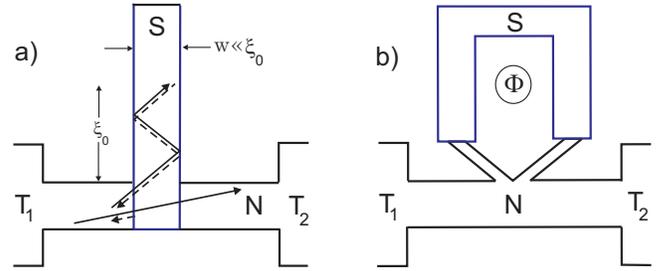}
\caption{(a) Upper view of the model of the NS interface in
Fig.~\ref{fig1}, with the ballistic trajectories of the electrons
(solid lines) and holes (dashed lines); (b) the geometry of
house-like Andreev interferometer with the magnetic flux $\Phi$.}
\label{fig2}\vspace{-5mm}
\end{figure}

Given a very satisfactory fit, the above line of rea\-son\-ing would
have been conclusive, unless the disturbing observation that in the
experiments of \cite{Chandr2}, the cross size $w$ of the N- and
S-leads was considerably smaller than $\xi_0$, making the Andreev
barrier semi-transparent for the subgap particles. This raises
serious doubts about the applicability of the above `macroscopic'
consideration, requiring $w\gg\xi_0$, to the quantum problem, $w <
\xi_0$, where the subgap quasiparticles may well dominate the thermal
conductance at low temperatures. This is the case in ballistic
structures, with the elastic scattering length $\ell$ much larger
than $\xi_0$ and $w$, where the heat flow is carried by
quasiparticles traversing the proximity region and overcoming the
weak Andreev barrier (see Fig.~\ref{fig2},a). Such effect was
considered within a model of a small S-inclusion in the ballistic
N-wire \cite{CL}. In the geometry of Fig.~\ref{fig2},a, the
quasiparticles which propagate into the S-lead over the distance
$\xi_0$, undergo complete Andreev reflection, and sink out from the
heat transport \cite{foot1}. In the limit of almost transparent
Andreev barrier, $w \ll \xi_0$, the ratio $\kappa/\kappa_N$ in the
ballistic structure is basically determined by the interface geometry
and therefore should be temperature independent, except in a vicinity
of $T_c$, where, as $\xi_0$ becomes larger than the length of the
S-lead, the Andreev reflection is suppressed, and $\kappa$ returns to
$\kappa_N$.

In this Letter, we focus on the diffusive limit, $\ell \ll
(w,\xi_0)$, that describes the experimental situation in Refs.\
\onlinecite{Chandr1,Chandr2} and reveals a wealth of interesting
physics. In this case, due to multiple coherent backscattering of
electrons by the impurities within the proximity region \cite{Wees},
the number of Andreev reflected quasiparticles effectively enhances.
This enhancement becomes especially pronounced as the energy
approaches zero, and, correspondingly, the size $\xi_N(E) =
\sqrt{\hbar{\cal D}/2E}$ of the proximity region in a normal metal
with the diffusion coefficient ${\cal D}$ infinitely increases. This
effect is known as the reason for the zero-bias conductance peak in
NS structures with an opaque interface \cite{VK}. In our context,
enhanced Andreev reflection means the suppression of the heat flow at
small energies even for the weak Andreev barrier, $w \ll \xi_0$, and,
as we show below, results in a rapid decrease of $\kappa/\kappa_N$
with temperature as observed in the experiment. Furthermore, the
temperature dependence of $\kappa$ appears to be power-like rather
than the exponential, in contrast to the model of the
``impenetrable'' Andreev barrier \cite{Chandr2,And} which blocks the
heat transport within the entire subgap region. Due to high
sensitivity of the heat flow to the proximity effect, one can also
expect essential dependence of the thermal conductance on the order
parameter phase difference $\phi$ applied to the proximity structure
(in the ballistic case, such dependence was predicted in
Ref.~\onlinecite{CL}). A hallmark of the phase-coherent heat
transport was found in thermopower measurements \cite{Chandr1,TP}.

We consider both electric and thermal conductances of the proximity
structure in Fig.~\ref{fig2},a in a diffusive limit, assuming the
N-lead of length $L_N$ ($-L_N/2<x<L_N/2$) to be connected to normal
reservoirs, and the S-lead of length $L_S$ ($0<y<L_S$) attached to
the middle of the N-lead. In such geometry, the quasiparticle flow
along the S-lead is blocked, and therefore the quasiparticle
distribution in this lead is spatially uniform. For this reason, it
is enough to solve one-dimensional diffusive kinetic
equations\cite{VK,LO} for the distribution functions $f_\pm$ in the
N-lead, $\partial_x \left( D_\pm\partial_x f_\pm\right) =  0$,
neglecting spatial variations across the leads at $w \ll \xi_0$. The
diffusion coefficients $D_\pm$ are defined through the retarded
branch of the spectral angle $\theta$ as $D_+ = \cos^2 \,\text{Im}\,
\theta$ and $D_- = \cosh^2 \,\text{Re}\, \theta$. The spectral angles
$\theta_N$ and $\theta_S$ in the N- and S-lead, respectively, are to
be determined from the Usadel equation
\begin{equation} \label{EqTheta}
2(E \sinh\theta - \Delta \cosh \theta) = i\hbar{\cal
D}\partial^2\theta,
\end{equation}
with the boundary conditions $\theta_N = 0$ at the normal reservoirs,
and $\partial_y\theta_S = 0$ at the edge of the S-lead. At the
transparent NS interface, the spectral angle is continuous,
$\theta_N(0) = \theta_S(0) \equiv \theta_0$, and obeys the current
conservation law \cite{Naz}, $g_S
\partial_y \theta_S(0) = \pm 2g_N \partial_x \theta_N(\mp 0)$, where
$g_{N,S}$ are the conductances of the leads per unit length.

The kinetic equations have the first integral,
\begin{equation} \label{f+}
D_\pm\partial_x f_\pm =  I_\pm(E),
\end{equation}
where the spatial constants $I_\pm$, which have the meaning of the
spectral densities of the probability and electric currents,
respectively, are to be determined from the boundary conditions for
$f_\pm$. The electric current $I$ and the thermal flow $Q$ in the
N-lead are related to $I_\pm$ as
\begin{equation} \label{IQ}
I = {g_N \over e} \int_0^\infty I_-(E)\,dE,\quad Q = {g_N \over e^2}
\int_0^\infty EI_+(E)\,dE.
\end{equation}

The thermal flow arises at different temperatures $T_1$ and $T_2$ of
the N-reservoirs, which imposes the boundary conditions for the
quasiparticle density function $f_+(\mp L_N/2) = \tanh (E/
2T_{1,2})$, where a small ther\-mo\-power is neglected. Assuming the
temperature difference to be small, $T_1 -T_2 \ll T = (T_1+T_2)/2$,
Eqs.~(\ref{f+}) and (\ref{IQ}) give the expression for the thermal
conductance,
\begin{equation} \label{kappa}
\kappa = {Q \over L_N(T_1-T_2)}= {3\kappa_N \over 2\pi^2 T^3}
\int_0^\infty {E^2 R_Q^{-1}(E)\, dE \over \cosh^2 (E / 2T)} ,
\end{equation}
where $R_Q(E)= \langle D_+^{-1}(E,x)\rangle $ is the dimensionless
spectral thermal resistance, and the angle brackets denote average
over the length of the N-lead.

In the problem of charge transport, we assume the left and right
reservoirs to be voltage biased at $\mp V/2$ and maintained at equal
temperature $T$, which results in the boundary condition for the
charge imbalance function $f_-$,
\begin{equation} \label{f-}
f_-(\pm L_N/2) = \!{1\over 2}\!\left[\tanh {E \!\pm\! eV/2 \over 2T}
- \tanh {E\! \mp\! eV/2 \over 2T} \right]\!\!.
\end{equation}
Using Eqs.~(\ref{f+}), (\ref{IQ}) and (\ref{f-}), we obtain the
zero-bias conductance of the N-wire,
\begin{equation} \label{G}
G =\left.{dI \over dV}\right|_{V \rightarrow 0} = {G_N\over
2T}\int_0^\infty { R_I^{-1}(E)\, dE \over \cosh^2 (E / 2T)} ,
\end{equation}
where $G_N=g_N/L_N$ is the normal conductance, and $R_I(E)= \langle
D_-^{-1}(E,x)\rangle $ is the spectral electric resistance.

\begin{figure}[tb]
\epsfxsize=8.5cm\epsffile{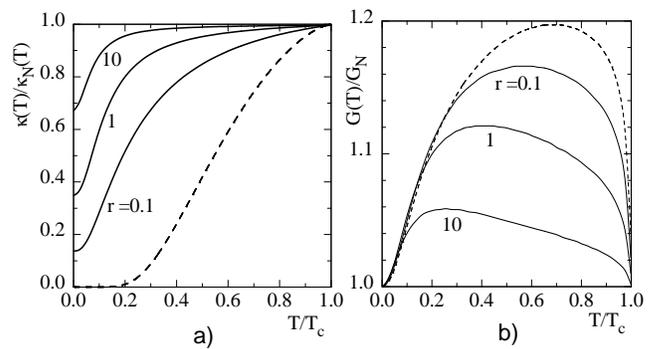} \vspace{0mm}
\caption{ Temperature dependencies of $\kappa/\kappa_N$ (a) and
$G/G_N$ (b) obtained by numerical solution of Eq.~(\ref{EqTheta}), in
comparison with the Andreev model (dashed curves), at $L_N=10\xi_0$.
} \label{fig3} \vspace{-5mm}
\end{figure}

The calculation of the spectral angle $\theta_N(E,x)$, which enters
the coefficients $D_\pm$, is to be performed numerically. However,
neglecting spatial variations in $\Delta$ within the proximity region
(that may result in some non-crucial numerical factors at most) and
assuming the lengths of the leads to be much larger than the
characteristic scales $\xi_{N,S}$ of spatial variation of the
spectral angle,
\begin{equation} \label{xi}
L_N \gg \xi_N = \sqrt{\hbar {\cal D}_N \over 2E}, \;\;  L_S \gg\xi_S
= \sqrt{\hbar {\cal D}_S \over 2\sqrt{\Delta^2-E^2}},
\end{equation}
one can apply the following analytical solutions of
Eq.~(\ref{EqTheta}) for the structure with semi-infinite leads,
\begin{eqnarray}
&\displaystyle \tanh {\theta_N(E,x) \over 4} = \tanh {\theta_0(E)
\over 4} e^{-|x|/ \xi_N \sqrt{i}},\label{thetaN}
\\
&\displaystyle \tanh {\theta_S(E,x)-\theta_{B} \over 4} = \tanh
{\theta_0(E)-\theta_B \over 4}e^{-y / \xi_S }, \label{thetaS}
\\
&\displaystyle \theta_0 = \ln\! { \sqrt{E+\Delta} + \!\sqrt{Er} \over
\sqrt{E-\Delta} + \!\sqrt{Er} }, \;\; r = {4{\cal D}_S \over {\cal
D}_N}\!\left( {g_N \over g_S} \right)^2 \!\! \sim \!\!{{\cal D}_N
\over {\cal D}_S}. \label{theta0}
\end{eqnarray}
Here $\theta_{B} = \,\text{Arctanh}\,(\Delta/E)$ is the spectral
angle in a bulk superconductor, and the parameter $r$ determines the
strength of the proximity effect (see comments to Fig.~\ref{fig3}).
Similar solution for the normal part of a long SNS structure was used
in Ref.\ \onlinecite{Zaikin}.

The semi-infinite-lead approximation enables us to estimate the
magnitude of $\kappa$ in the most interesting case of small
temperatures, $T \ll \Delta$. Although the coefficient $D_+$ is
suppressed within the whole proximity region, $|x| \lesssim \xi_N\sim
\xi_0 \sqrt{\Delta/T}$, the contribution $\delta R_Q$ of this region
to the thermal resistance $R_Q$ comes from the narrow (of the order
of $\xi_0 = \sqrt{\hbar {\cal D}_N/2\Delta}$) vicinity of the
crossing point, in a contrast to the proximity correction to the
zero-bias electric conductance which is formed over the much larger
scale $\xi_N$. Within this region, the coefficient $D_+$ is
anomalously small and turns to zero at $E \rightarrow 0$,
\begin{equation} \label{Dapp}
D_+ \approx  (E / \Delta) \left(|x|/\xi_0 + \sqrt{r}\right)^2,
\end{equation}
which results in the following estimate of $\delta R_Q$ at the
characteristic energies $E \sim T$,
\begin{equation} \label{RTapp}
\delta R_Q(T) \approx {T_0 / T}, \quad T_0 = \Delta {\xi_0 /
L_N\sqrt{r}}.
\end{equation}
Thus, at $T < T_0$, the proximity region dominates the net thermal
resistance giving rise to the power-like decrease of $\kappa \sim
\kappa_N T/T_0 \sim T^4$ with $T$. At very low temperatures, smaller
than the Tho\-u\-less energy $E_{\text{Th}}= \hbar {\cal D}_N/L_N^2$,
the N-reservoirs begin to affect the quasiparticle spectrum at the
contact area, and the approximation of semi-infinite N-lead fails. In
this case, $E_{\text{Th}}$ provides the cutoff in the decrease of the
coefficient $D_+$ at small $E$, which therefore saturates at $D_+
\sim E_{\text{Th}}/\Delta$. As the result, the thermal conductance at
$T < E_{\rm Th}$ starts to decrease slower, as $\kappa_N \sim T^3$,
with a small prefactor $E_{\text{Th}}/T_0\sim \xi_0 \sqrt{r}/L_N$. At
the same time, the electric conductance shows behavior typical for
the proximity structures: as $T$ decreases, $G(T)$ approaches maximum
and then returns to $G_N$.

Shown in Fig.~\ref{fig3} are the results of calculation of thermal
and electric conductances within the whole relevant temperature
region $0<T<T_c$ for different mag\-nitudes of the parameter $r$. If
the conductivity of the N-lead is small, $r \ll 1$, the spectrum
within the S-lead is almost unperturbed [$\theta_0 \approx
\theta_{B}$, see Eq.~(\ref{theta0})], and the crossing area plays the
role of a su\-per\-con\-duc\-ting reservoir. In terms of
quasiparticle motion, quasiparticles `would prefer' to diffuse into
the high-con\-duc\-tive S-lead, where most of them undergo Andreev
reflection thus additionally reducing $\kappa$ and increasing $G$. In
the opposite limit, $r \gg 1$, the quasiparticles avoid penetration
into the low-conductive S-lead, which suppresses the proximity effect
($\theta_0 \to 0$), and, correspondingly, enhances $\kappa$ and
reduces the peak in $G(T)$. Note that for all reasonable values of
$r$, our quantum model gives the magnitude of $\kappa$ considerably
larger than that due to the classic model \cite{Chandr2,And}.

The high sensitivity of thermal conductance to the details of the
quasiparticle spectrum can be used for a study of the effects of
phase coherence in the mesoscopic devices known as Andreev
interferometers. Usually, the interferometer circuits contain a
superconducting loop with the order parameter phase difference $\phi$
controlled, e.g., by the magnetic flux, and connected to an SNS
junction.  The electric conductance of the junction (or the N-wire
attached to the junction) depends periodically on $\phi$.

The basic features of the phase-coherent heat transport in the
proximity structures can be demonstrated within the model of the
``house interferometer'' \cite{Chandr1,Chandr2} shown in
Fig.~\ref{fig2},b. Assuming the length of the junction arms to be
smaller than $\xi_0$, and the normal conductivity of the S-loop to be
much larger than the conductivity of the N-wire, one can neglect the
change in the junction quasiparticle spectrum due to proximity to the
N-wire. In this limit, the spectrum at the contact area is similar to
the BCS-like spectrum of a separate short diffusive SNS junction,
with the phase-dependent energy gap $E_g(\phi) = \Delta |\cos
(\phi/2)|$ \cite{KO}. This implies that the contact area behaves as
an impenetrable Andreev barrier with the phase-dependent height
$E_g(\phi)$, and therefore the heat transfer is governed by the
quasiparticles with the energies $E > E_g(\phi)$. Within this region,
we apply Eq.~(\ref{kappa}), thus taking into account spatial
variations in the diffusion coefficient $D_+$ within the proximity
region. These variations are a diffusive analog \cite{circuit} of the
ballistic over-the-bar\-ri\-er Andreev reflection from a rapidly
varying order parameter potential.

\begin{figure}[tb]
\epsfxsize=8.5cm\epsffile{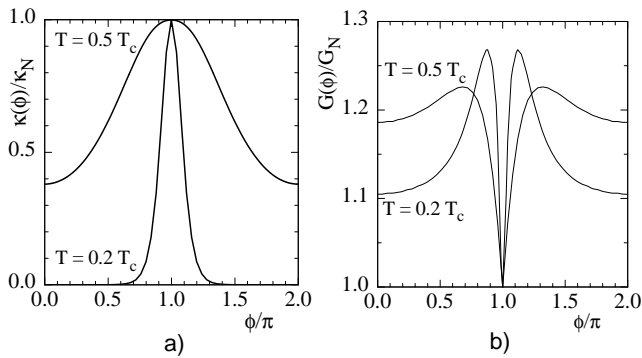}
\caption{ Phase dependencies of the thermal (a) and electric (b)
conductances in the geometry of house interferometer in
Fig.\ref{fig2},b, for different temperatures ($L_N=10\xi_0)$. }
\label{fig4} \vspace{-5mm}
\end{figure}

The phase dependence of the thermal conductance is presented in
Fig.~\ref{fig4},a. Due to Andreev reflection from the contact area,
$\kappa$ is generally suppressed, according to the mechanism
discussed above. However, as the phase difference approaches odd
multiples of $\pi$, the energy gap $E_g$ closes, and all
quasiparticles freely diffuse through the N-wire, which restores the
normal value of $\kappa$. Thus, at low temperatures, $\kappa(\phi)$
exhibits strong oscillations with the phase $\phi$, with sharp peaks
at $\phi\, \text{mod} \,2\pi = \pi$. The electric conductance shows
oscillations with sharp dips at the same points, but with much lower
amplitude (Fig.~\ref{fig4},b).

In conclusion, we have developed a theory of the heat transport
through the normal part of diffusive NS na\-no\-struc\-tures. The
Andreev reflection from the NS interface blocks the quasiparticle
probability current, and therefore the heat flow through the normal
wire, coupled with the superconductor over a small area, is
essentially suppressed. Our approach takes into account
self-con\-sis\-ten\-t\-ly the influence of both the normal metal and
superconductor on the quasiparticle spectrum within the proximity
region, extending thus the ballistic concept of partial Andreev
reflection from the superconductor of finite width to diffusive
proximity systems. The important feature of the diffusive structures
is the essential enhancement of the Andreev reflection probability at
low energies due to multiple returns of coherent quasiparticles to
the contact area. For this reason, the proximity region dominates the
net thermal resistance of a long normal wire with the length $L_N \gg
\xi_0$ at low temperatures, even if the size of the contact area is
much smaller than the coherence length $\xi_0$. This results in a
power-law decrease in the thermal conductance with the temperature,
$\kappa \sim T^4$, transforming into a cubic law at the temperatures
smaller than the Thouless energy. The effect becomes more pronounced
when the carriers mobility in the superconductor is higher than in
the normal metal. If the wire is at\-tached to the SNS junction with
the order parameter phase difference $\phi$ between the electrodes,
the thermal conductance reveals full-scale oscillations with $\phi$
and shows large peaks at odd multiples of $\pi$.

In the experiments \cite{Chandr2}, the results of measurements of
$\kappa(T)$ were fitted by the Andreev formula \cite{And}, which
assumes complete blockade of the thermal flow within the entire
subgap region, $E < \Delta$, and therefore leads to the exponential
temperature dependence of $\kappa$. Although this fitting looks
satisfactory, the physical background for the applicability of such a
simple model remains unclear, because of the small width $w \approx
0.4\xi_0$ of the wires and unavoidable suppression of electron-hole
correlations in the S-wire within the proximity region. This calls
for further investigations, including independent measurements of
$\kappa(T)$ in the normal state and the extension of theoretical
calculations for rather complicated geometry of real proximity
structure in Ref.\ \onlinecite{Chandr2}.

It is a pleasure to thank V.~Chandrasekhar, Yu.~M.~Gal\-pe\-rin,
V.~I.~ Kozub, A.~N.~Shelankov, and V.~S.~Shu\-mei\-ko for fruitful
discussions. This work was supported by the U.S. Department of
Energy, Office of Science under contract No. W-31-109-ENG-38.


\begin{thebibliography}{99}
\bibitem{foot1} We note that Andreev reflection, due to its quantum nature,
develops continuously at the scale of $\xi_0$, and therefore the
picture of instantaneous electron-hole conversion in
Fig.~\ref{fig2},a is rather conditional.
\end{thebibliography}
\end{document}